\documentclass[twoside]{article}

\usepackage{lipsum} % Package to generate dummy text throughout this template
\usepackage{aas_macros}

\usepackage[sc]{mathpazo} % Use the Palatino font
\usepackage[T1]{fontenc} % Use 8-bit encoding that has 256 glyphs
\usepackage{microtype} % Slightly tweak font spacing for aesthetics

\usepackage[hmarginratio=1:1,top=32mm,columnsep=20pt]{geometry} % Document margins
\usepackage{multicol} % Used for the two-column layout of the document
\usepackage[hang, small,labelfont=bf,up,textfont=it,up]{caption} % Custom captions under/above floats in tables or figures
\usepackage{booktabs} % Horizontal rules in tables
\usepackage{float} % Required for tables and figures in the multi-column environment - they need to be placed in specific locations with the [H] (e.g. \begin{table}[H])
\usepackage{hyperref} % For hyperlinks in the PDF
\usepackage{natbib} %- my add
\usepackage{graphicx}
\usepackage{caption} %subfigures - my add
\usepackage{subcaption} %subfigures - my add
\usepackage{lettrine} % The lettrine is the first enlarged letter at the beginning of the text
\usepackage{paralist} % Used for the compactitem environment which makes bullet points with less space between them
\usepackage{wasysym}
\usepackage{stmaryrd}

\usepackage{abstract} % Allows abstract customization
\usepackage{color}

\usepackage{ulem}

\usepackage{titlesec} % Allows customization of titles
\renewcommand\thesection{\Roman{section}} % Roman numerals for the sections
\renewcommand\thesubsection{\thesection.\arabic{subsection}} % Roman numeralsfor subsections
\titleformat{\section}[block]{\large\scshape\centering}{\thesection.}{1em}{} % Change the look of the section titles
\titleformat{\subsection}[block]{\large}{\thesubsection.}{1em}{} % Change the look of the section titles

\usepackage{fancyhdr} % Headers and footers
\title{\vspace{-15mm}\fontsize{24pt}{10pt}\selectfont\textbf{The Existential Threat of Future Exoplanet Discoveries}} % Article title

\author{
\large
\textsc{Michael B. Lund$^1$}\\%, Robert J. Siverd$^2$, and Ponder Stibbons$^3$}\\%\thanks{A thank you or further information}\\[2mm] % Your name
\normalsize $^1$CalTech/IPAC-NExScI\\
\normalsize \href{mailto:mlund@ipac.caltech.edu}{mlund@ipac.caltech.edu} % Your email address
\vspace{-5mm}
}
\date{}

\begin{document}

\maketitle % Insert title

\thispagestyle{fancy} % All pages have headers and footers

%----------------------------------------------------------------------------------------
%	ABSTRACT
%----------------------------------------------------------------------------------------

\begin{abstract}

\noindent The last 25 years have been revolutionary in astronomy, as the field of exoplanets has gone from no known planets outside the Solar System to thousands discovered over the last year few years. This represents a rapid increase not just in known planets (often referred to as Mamajek's Law), but also in total planetary mass. What has been heretofore unaddressed, however, is that this rapid increase in planetary masses may have disastrous consequences for the future of humanity. We look at how the number of planets, and more importantly, the mass of these planets has changed in the past and how we can expect this to change in the future. The answers to those questions, and how we respond to them, will determine if humanity is able to survive beyond the next 230 years.

\end{abstract}

%----------------------------------------------------------------------------------------
%	ARTICLE CONTENTS
%----------------------------------------------------------------------------------------

\begin{multicols}{2} % Two-column layout throughout the main article text

\section{Introduction}
\begin{figure*}[!htb]
  \begin{center}
   \includegraphics[width=\textwidth]{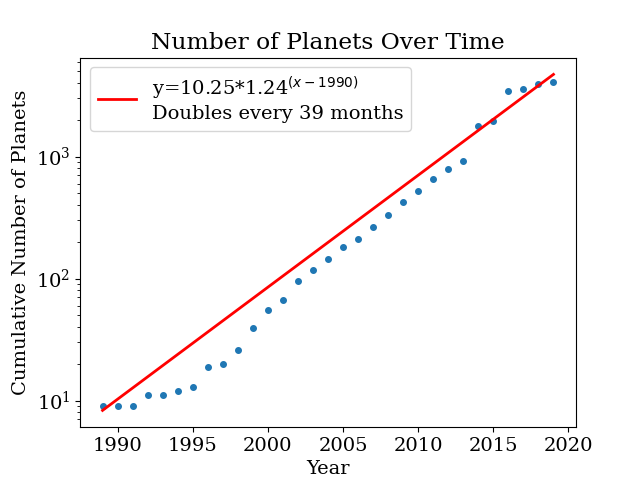}
  \end{center}
  \caption{The number of planets as a function of time, with data points in blue and the best-fit exponential in red. This is similar to that proposed in Mamajek's Law but includes planets within our own solar system as well. The number of planets doubles every 39 months.}
  \label{fig:Mamajek_Law}
\end{figure*}

\lettrine[nindent=0em,lines=3]{F}or hundreds of years, the only planets that were discovered were those within our own solar system \citep{Herschel1781, Galle1846}. By the early 1990s, however, the first planets started to be officially discovered and confirmed outside our solar system \citep{Wolszczan1992, Mayor1995}. The confirmation of these exoplanets led to a significant increase of interest in the field and optimistic expectations of the forthcoming yields of exoplanets from new missions designed to find these objects \citep{Horne2003}. The rate of exoplanets discoveries has continued at an ever-quickening pace and has been referred to as Mamajek's Law\footnote{https://twitter.com/EricMamajek/status/790767025580707840}, similar to the fashion in which Moore's law describes the complexity of integrated circuits as doubling roughly every year \citep{Moore1975}. In Figure~\ref{fig:Mamajek_Law}, we provide a current calculation of a slightly modified version of Mamajek's Law, including the known planets within the solar system as well and using the data from the NASA Exoplanet Archive\footnote{https://exoplanetarchive.ipac.caltech.edu/}. From the first exoplanets discovered through 2019, the number of known planets has been doubling roughly every 39 months, and the cumulative number of exoplanets ($N_{\textrm{planets}}$) can be represented by the following equation:
\begin{equation}
N_{planets} = 10.25 * 1.24^{(year - 1990)}
\end{equation}

While in the short term the implications of Moore's Law and Mamajek's Law simply suggest that we will see much faster computing power and many more exoplanets in the near future, respectively, the long term consequences can be more serious. This has largely been ignored over the last few decades in astronomy, and it has been noted that scientists were so preoccupied with whether or not they could, they didn't stop to think if they should \citep{Malcolm1993}. A future consequence of the rapid increase in computing capabilities has been discussed in the last few decadse and is known as the impending Technological Singularity, such a rapid breakthrough in technology that it will mark the end of the human era on Earth \citep{Vinge1993}. Astronomy, in contrast, has been familiar with the consequences of singularities for a somewhat longer period of time \citep{Schwarzschild1916, Chandrasekhar1931} but has not applied this concept to the growth of exoplanets.

In this work we extend the consequences of the increasing rate of new planets beyond the immediate future and consider the total mass of known exoplanets. Specifically, we address how long this rate of new exoplanets can last before the added mass from new planets is so great that it manifests as a black hole. This would potentially end not just humanity on Earth, but all life in our area of the universe. We present our methodology in Section~\ref{Methods} and discuss our results in Section~\ref{Discussion}, then we summarize this paper in Section~\ref{Summary}.

%------------------------------------------------
\section{Methods} \label{Methods}

\subsection{Data}
For gathering all exoplanet data, we use tables available at the NASA Exoplanet Archive supported by the NASA Exoplanet Science Institute, which compiles all published and confirmed exoplanets. As there is some delay between announcement and publishing that can take place on the order of months, we only look at planets that have a discovery year prior to 2020. For gathering exoplanet masses, we use the data available from the gamma release of the Planetary Systems Composite Parameters (PSCP) Table \citep{PSCT}. The value of this table is that in order to provide complete parameters, the Exoplanet Archive consolidates masses derived from several different approaches. While the most reliable are masses that have been directly calculated (either in a publication or directly by the Exoplanet Archive from published parameters), planets with unconstrained inclinations still provide a minimum mass, and planets with only a radius have masses calculated using a mass-radius relationship. In this way, we are able to have masses for a very heterogeneous population of exoplanets to carry out the rest of our calculations.

\subsection{Population Characteristics}
\begin{figure*}[!htb]
  \begin{center}
   \includegraphics[width=\textwidth]{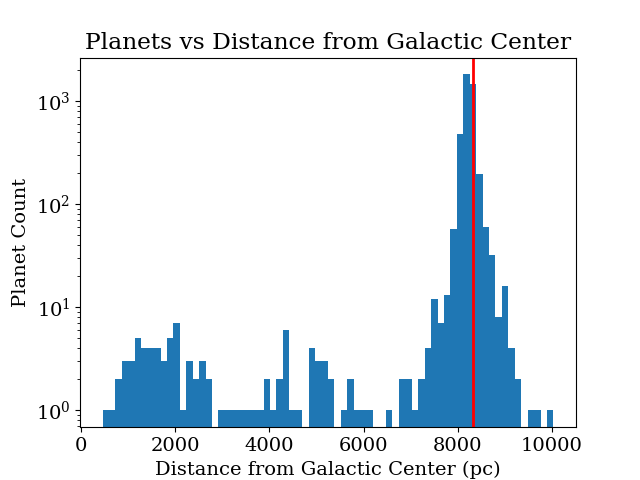}
  \end{center}
  \caption{Number of planets vs distance from the center of the galaxy. The red horizontal line marks the distance that the solar system is from the center of the galaxy. Prominent peaks represent microlensing planets close to the center of the galaxy on the left and planets closer to the earth near the earth's distance.}
  \label{fig:planets_distance}
\end{figure*}

Distances from Earth and galactic coordinates (in conjunction with 8.32 kpc as the distance between the Earth and the center of the galaxy \citep{Gillessen2017}) were used to determine how distant each planet is from the center of the galaxy, and is shown in Figure~\ref{fig:planets_distance}. There are two prominent peaks in planet counts as a function of distance from the galactic center, with the first peak being under 2,000 pc and largely representing microlensing planets that have been discovered near the galactic bulge. The more prominent peak is roughly around the distance that our solar system is from the center of the galaxy (shown by the red horizontal line), generally from transiting and radial velocity searches. The furthest planets are $\sim$ 10 kpc from the center of the galaxy.

\subsection{Calculations}

The Schwarzschild radius represents the radius of the event horizon for a black hole of a given mass, and can be expressed by the following equation:

\begin{equation}
    r_{s} = \frac{2 G M}{c^{2}}
\end{equation}

This same equation can be rewritten to solve for mass:

\begin{equation}
    M = \frac{r_{s} c^{2}}{2 G}
\end{equation}

As all exoplanets are within 10 kpc of the center of the galaxy, that Schwarzschild radius would correspond to a total mass of $1.05 \times 10^{17} M_{\odot}$. This is well above the currently known mass of the Milky Way within 10 kpc at $\sim0.3 \times 10^{12} M_{\odot}$ \citep{Kafle2014}, and so we can be reasonably certain that we are not, in fact, already inside of a black hole. As we are dealing with planetary and not stellar masses, it is more useful to think of the critical mass in Jovian masses as $1.09 \times 10^{20} M_{J}$ and the current mass within that 10 kpc radius as $\sim\pi \times 10^{14} M_{J}$.

In keeping with Mamajek's Law, our primary equation can be described as a natural exponential equation of the following form, where the best-fit values are a = 342 and b= $1.59 \times 10^{-1}$, corresponding to a doubling time of $\sim4.4$ years:  
\begin{equation}
    M = a \times e^{b(year-2000)}
\end{equation}

For the sake of completeness, we do also consider two other functions to fit to the data, although these do not fit the data as well currently. The first of these equations is a simple quadratic equation, where the best fit values are a = 13.2, b = $-5.26 \times 10^4$, and c = $5.25 \times 10^7$:
\begin{equation}
    M = a \times (year)^2 + b \times year + c
\end{equation}
The second additional equation we consider is a piecewise function of an exponential function that at some point turns into a straight line. This function does carry the further stipulation that it is continuous and differentiable. In this case the best values are a= $2.82 \times 10^3$, b = 0.22, and c = 2012.8, corresponding to a break in the functional behavior in October 2012:
\begin{equation}
    M =
    \left\{ \begin{array}{ll}
        a \times e^{b(year-c)} & year < c \\
        ab(year-c) + a & year \geq c
    \end{array} \right.
\end{equation}

\begin{figure*}[!htb]
  \begin{center}
   \includegraphics[width=\textwidth]{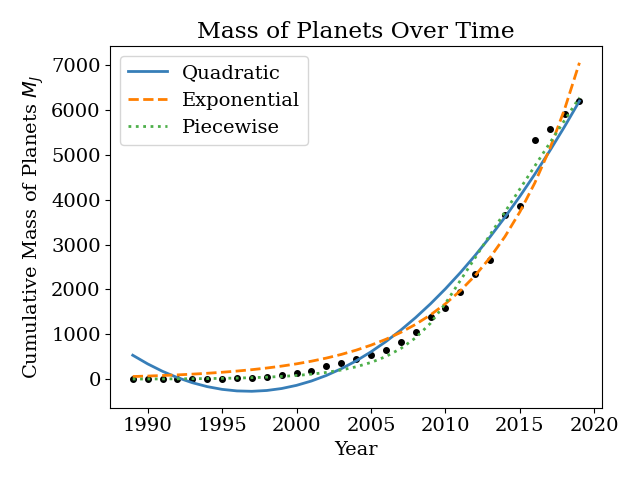}
  \end{center}
  \caption{Cumulative masses of planets as a function of year. Black points represent data from the NASA Exoplanet Archive. The blue solid line represents the best-fit quadratic function, the orange dashed line represents the best-fit exponential function, and the green dotted line is a piecewise function that combines an exponential function and a straight line.}
  \label{fig:planetmass_time}
\end{figure*}

We plot the cumulative mass of planets and these three functions in Figure~\ref{fig:planetmass_time}, with the data points in black, and the quadratic function (solid blue line), exponential function (dashed orange line), and piecewise function (dotted green line) all plotted over that.

We then determine in what year each of these functions will reach the critical mass of $1.09 \times 10^{20} M_{J}$. For the quadratic function, this occurs in 2.9 Gyr, and for the piecewise function this occurs in $1.8 \times 10^{17}$ years. We maintain our focus on the exponential function, however, for two primary reasons. Firstly, the quadratic function does suggest a negative planetary mass for much of the 1990s, and we believe that had such a loss of mass occurred in that decade, it would have been noticed. Secondly, the piecewise function, while ostensibly a better fit by virtue of no negative values, does not appear to have any physical justification for that break-point. We believe our precision is sufficient to rule out a break point of 2012.97, which would provide a rationale for a change in the function as it would correspond with the end of the Mayan calendar \citep{Larsen2010, Manning2010}.

With these functions ruled out, we are left with our initial exponential function that is similar to Mamajek's Law, with masses doubling every 52 months. In this case, our rate of planet discovery will reach the singularity late in the year of 2252. This year has already been focused on as noteworthy in some literature \citep{Haldeman2008}.

\section{Discussion}\label{Discussion}
Prior to this work, the rapid rate of exoplanet discoveries has almost exclusively been viewed as a positive series of events, providing ever richer data sets to further our understanding of planet formation and perhaps even address questions regarding the prevalence of life in the universe. These are, to be sure, valuable questions with important answers. There remains, however, a darker and more sinister side to this rapid increase in mass. The change from no known exoplanets to the thousands that have been discovered thus far has put us on a collision course with disaster, as we have only about 230 years until the mass within 10 kpc of the galactic center satisfies the conditions to create a singularity.

The greatest risk, perhaps, is most strongly in the uncertainty of what lies after that. The interior of a black hole has been speculated about, but still remains a mystery to outside observers \citep{BlackHole1979, Nolan2014}. We don't even know if this black hole sun will come and wash away the rain \citep{Soundgarden1994, Lund2020}. In many regards, we are at least fortunate that this effect is being only caused by exoplanets, as the relative paucity of brown dwarfs has meant that this is not also coupled to the rate of increase in the mass of known brown dwarfs. Indeed, it could well be argued that the only reason we have as much time to prepare as we do before reaching a galactic calamity is that the brown dwarf desert means that we are approaching disaster at a slower rate than we otherwise would be \citep{Marcy2000}.

\section{Summary}\label{Summary}
In this work we have extended the core concept underlying Mamajek's Law to look not just at the number of planets that have been discovered, but also the total masses of these planets, using the NASA Exoplanet Archive's data set on measured and calculated planetary masses. At the current rate, the number of planets is doubling roughly every 39 months, and the cumulative mass is doubling slightly slower at roughly every 52 months, likely reflecting a shift away from Hot Jupiters as the dominant exoplanets being found.

Effectively all exoplanets have been found within 10 kpc of the center of the Milky Way. At the exponential rate that new exoplanets are contributing mass, the total mass within that radius will exceed the necessary minimum mass to form a black hole in about 230 years. The full consequences of this are open for speculation due to the number of questions that remain regarding the nature of black holes, but just like those black holes, if we get too close to this fate it will become inescapable.

\section{Acknowledgements}
This research has made use of the NASA Exoplanet Archive, which is operated by the California Institute of Technology, under contract with the National Aeronautics and Space Administration under the Exoplanet Exploration Program.

This research made use of Astropy,\footnote{http://www.astropy.org} a community-developed core Python package for Astronomy \citep{astropy:2013, astropy:2018}.

This research has made use of NASA’s Astrophysics Data System.

Software: astropy \citep{astropy:2018}, matplotlib \citep{Hunter2007}, numpy \citep{Oliphant2006}, pandas \citep{Mckinney2011}, scipy\citep{Virtanen2020}

%----------------------------------------------------------------------------------------
%	REFERENCE LIST
%----------------------------------------------------------------------------------------

\bibliographystyle{apalike}
\bibliography{main}

%----------------------------------------------------------------------------------------

\end{multicols}

\end{document}